\documentclass[12pt]{iopart}

\usepackage{graphicx}
\begin{document}

\title{Investigation of the formation process of soap bubbles from soap film}

\author{Jun Su, Weiguo Wang, Peng Xu, Ding Gu and Yuhao Zhou}

\address{Haian Senior School of Jiangsu Province, Jiangsu Province, China}

\ead{jsunnu@163.com}
\vspace{10pt}
\begin{indented}
\item[]September 2017
\end{indented}

\begin{abstract}
In this paper, we study the formation process of a soap bubble by blowing soap film. Both bubble diameter and formation position were investigated in experiments. We found that the ratio between bubble size and soap film column is constant, and that the formation position increases linearly within a critical length-range. We used the theory of Rayleigh-Plateau instability to explain these findings. The theoretical explanations are consistent with the experimental data.
\end{abstract}

%
%
%
%
%

\section{Introduction}
Soap bubbles are fascinating for young and old and they have been studied for centuries\cite{1,2}. When air from a nozzle blows at a soap-film\cite{3}, provided the dynamic pressure of the airflow $\sqrt {8\sigma /\rho {R_N}}$ is equal to the pressure difference ${\rm{4}}\sigma {\rm{/}}{R_N}$ (the airflow velocity just produces the bubble can define a critical value ${v_c} = \sqrt {\frac{{8\sigma }}{{\rho {R_N}}}}$), a bubble is successfully formed. Here, ${R_N}$ denotes the nozzle radius, $\rho$ is the air density, $\sigma$ is the surface tension between air and soap-water solution, and ${v_N}$ is the average airflow velocity at the nozzle.

What does the bubble-formation mechanism look like when the airflow velocity exceeds ${v_c}$? It is known that the length of a liquid jet increases when it falls vertically under gravity. However, the jet loses its shape due to hydrodynamic instability resulting from surface tension, and disintegrates into a stream of tiny droplets. This droplet formation is known as Rayleigh breakup, and was described by both Plateau and Lord Rayleigh already more than a century ago\cite{1,5}. The Rayleigh-Plateau instability is an ubiquitous phenomenon that occurs, for example, in water flowing from a kitchen faucet or any emulsification process\cite{6,7}. There are many interesting works of soap bubbles, the physics of soap bubble is illustrated in painting and art\cite{one06}, the relation between the radius of soap bubble and the internal pressure can be demonstrated in classroom\cite{one07}, and the internal pressure of soap bubble is also measured to determine the surface tension\cite{one08}. The process of blowing soap bubble is also study by \cite{8x}, the relation between the minimal radius of the thread and the time remaining till pinch-off can be represented by one scaling law, the scaling exponent of soap bubble pinch-off is about 2/3.

In this work, we conducted experiments to investigate both soap-bubble size and the formation position of the bubble, using different airflow velocities and different nozzle sizes. The obtained experimental data can be explained using Rayleigh-Plateau instability theory. The remainder of the paper is organized as follows: In section 2, we describe the experimental setup in detail, as well as the size of the soap bubble and the formation position studied. In section 3, we briefly discuss the Rayleigh-Plateau instability for a soap-film column, and we introduce the concept of the dispersion relation. We also discuss the ratio between the size of the bubble and the length of the soap film column. A linear relation between the formation position within a critical length range was found. Section 4 focuses on the analysis of the data. The results agree well with theoretical prediction.
\section{Experimental Setup}
We used four sizes of nozzles, which were manufactured by a 3D printer. Their inner diameters were 0.50 cm, 0.75 cm, 1.20 cm, and 1.70 cm. A soap-water solution with a surface-tension coefficient $\sigma$ of 0.038 N/m. The surface-tension coefficient was measured using a glass capillary tube with an inner diameter of 1.5 mm. The solution was a mixture of tap water and commercial soap, and the soap concentration was 2.6\% by volume. A hot-wire anemometer (CEM DT-8880) was used to measure the average speed at the nozzle, the size of the sensor of anemometer is comparable to nozzle's diameter, so we did't considered the distribution of velocity at the nozzle. The measurement range was 0 $\sim$ 30 m/s and the accuracy 0.01 m/s. A triangular plastic frame (the length of one side is about 20 cm) was used for the generation of soap film. The size of the nozzles was much smaller than the soap film. We used a mini fan connected with the nozzle to produce the air jet and let the jet blow the soap film. A high-speed camera (SONY RX100IV), capable of recording 500 frames per second, was used to record the bubble formation process, viewed from the side. During the experiments, the nozzles were positioned close to the liquid film, and the airflow velocity at the nozzle was adjusted by changing the voltage of the power supply driving the fan. The videos were analyzed using the open source software, Tracker\cite{8}. Several important physical parameters, including the bubble formation position (the distance between the nozzle and the breakup point), and the diameters of film column and bubble, were measured.

\begin{figure}
    \begin{center}
        \includegraphics[height=0.3\linewidth]{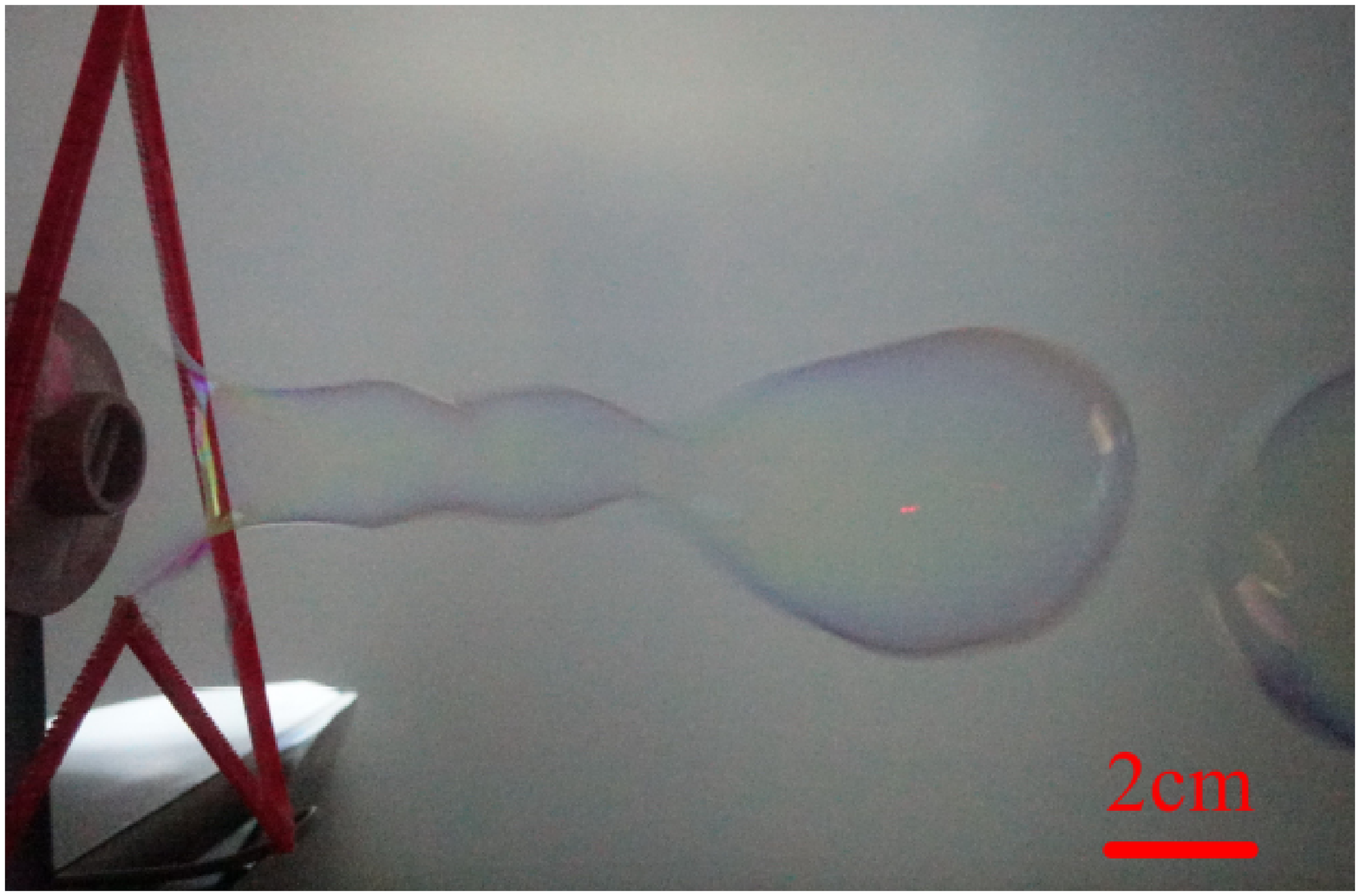}
        \includegraphics[height=0.3\linewidth]{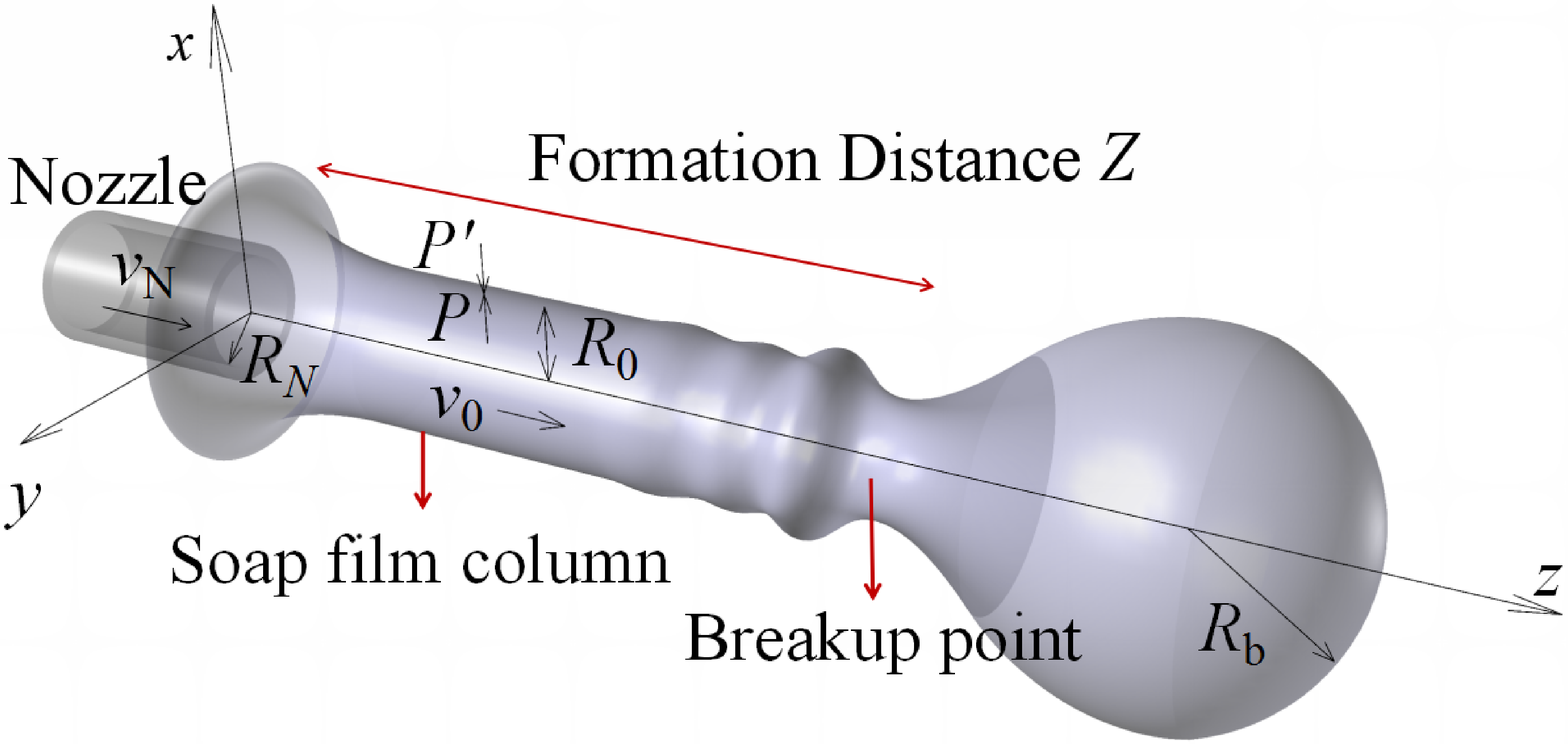}
    \end{center}
\caption{Left: The image of soap bubble's formation by blowing soap film. Right: The schematic of soap bubble's formation process.}
\end{figure}

The left in figure 1 shows an image of the bubble formation when blowing a soap film. The right in figure 1 shows a schematic of the soap-bubble formation process when the air jet from the nozzle produces the soap film. The axis of the jet is parallel to the $z$ axis. A cylindrical shaped hollow part is formed first, which we call the soap film column. The small perturbation on the surface of film column increases quickly before the film column is pinched off. ${R_N}$ denotes the radius of nozzle and ${R_0}$ is the radius of soap film column. We noticed that the generated soap film column is generally larger than the nozzle diameter in all experiments. Furthermore, $Z$ is the formation position of the bubble, ${R_b}$ is the radius of the bubble, $\rho$ and $P$ are the air density and pressure inside the film column, ${\rho _0}$ and ${P_0}$ are the air density and pressure outside the film column, respectively; ${v_N}$ is the air velocity at the nozzle, $v_0$ is the air velocity inside the film column. We suppose that the mass for air  in the soap film column is conserved, and neglect the effect of airflow drawn into the film column through the gap between the nozzle and the soap film, we have ${v_N}R_N^2 = {v_0}R_0^2$. The bubbles were initially non-spherical due to vibration after bubble formation. However, the shape of the bubbles stabilize and become spherical quickly.

\begin{figure}
    \begin{center}
        \includegraphics[height=0.32\linewidth]{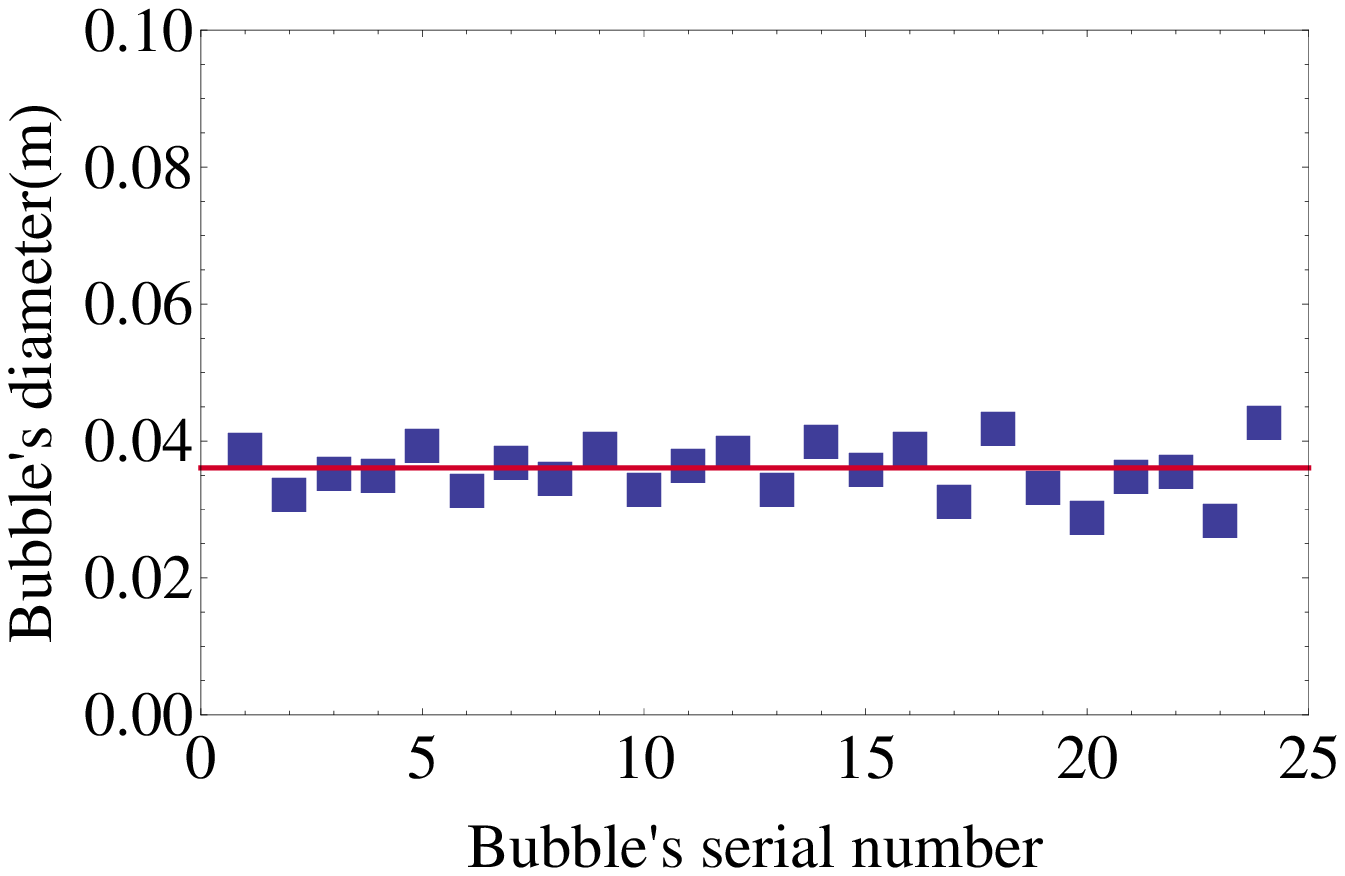}
        \includegraphics[height=0.32\linewidth]{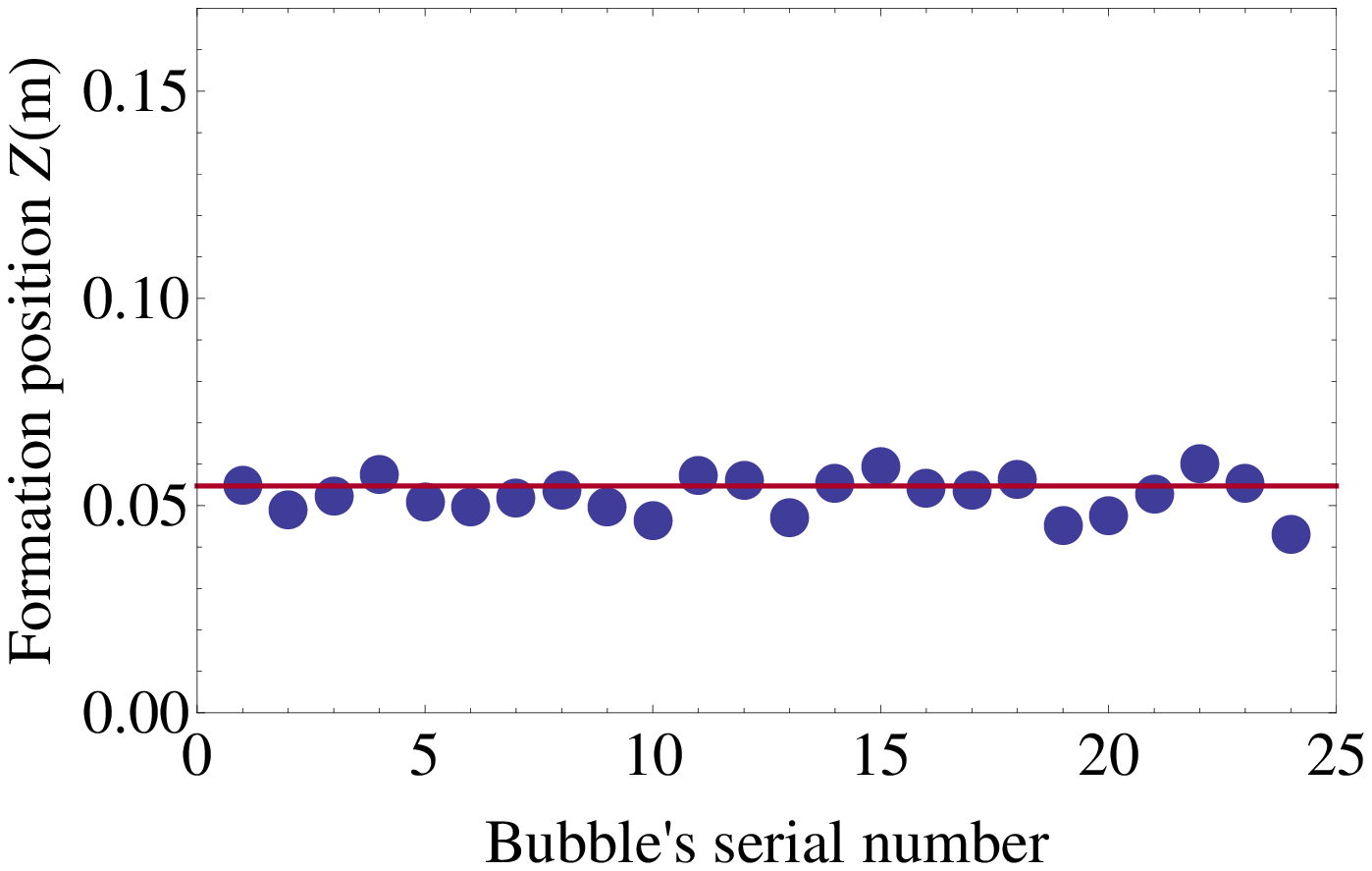}
    \end{center}
\caption{Left: Bubble's diameters for the 0.75 cm diameter nozzle at airflow velocity of 8.71 m/s. Right: Bubble's formation position for the 0.75 cm diameter nozzle at airflow velocity of 8.71 m/s. The solid red line is the optimal fitted value with a constant.}
\label{fig2}
\end{figure}

We conducted pre-experiments with the nozzles to identify the minimum and maximum blowing velocity needed for the bubble formation. When the airflow velocity is below a critical value ${v_c}$, the film bulges but no bubble is formed. When the airflow exceeds ${v_c}$, a film column forms and breaks up at a certain position to produce a bubble. When the airflow exceeds the critical airflow ${v_{\max }}$, the liquid film fractures, and no bubble is formed. Therefore, in subsequent experiments, the airflow was always within this speed range. The soap-water solution had a density of ${\rho _b} = {10^3}{\rm{kg/}}{{\rm{m}}^{\rm{3}}}$.

When the film thickness $h$ was about tens of micrometers\cite{9}, the bubble radius was about 2.5 cm. We estimate the Bond number using $Bo = \frac{{{\rho _b}{R_b}hg}}{{2\sigma }} < 0.036$, which indicates that gravity plays a negligible role here. Using $\mu  = 1.8 \times {10^{ - 5}}{\rm{Pa}} \cdot {\rm{S}}$ and $\rho  = 1.2{\rm{kg/}}{{\rm{m}}^{\rm{3}}}$ for the viscosity and density of air, respectively, we estimate the Ohnesorge number as $Oh = \frac{\mu }{{\sqrt {2\sigma {R_b}\rho } }} < 0.0004$. Effects due to the viscosity of air are also negligible.

We studied the 0.75 cm diameter nozzle first. The velocity range for this nozzle is about 8.52 m/s $\sim$ 22.60 m/s. We investigated the bubble size for five different velocities: 8.71 m/s, 9.70 m/s, 13.14 m/s, 16.85 m/s, and 20.60 m/s. The bubble diameters were measured for five different velocities. The bubble diameters at 8.71 m/s airflow are shown on the left in figure 2. We can see that the bubble sizes were roughly uniform, and a constant was used to fit the data. The fitted value was 0.036 m. The same method was used for the other four airflow velocities. Using the five fitted diameter values, we investigated the relationship between the bubble radius ${R_b}$ and airflow velocity ${v_N}$. The radius of the film column ${R_0}$ varied for different air flows ${v_N}$, therefore, the corresponding ${R_0}$ for each air flow ${v_N}$ was measured using the software. We plotted ${R_b}/{R_0}$ versus ${v_N}$ in figure 4. The bubble size remained nearly unchanged as the airflow increased, see figure 4.

We used the same method to study the bubble formation positions for the five different velocities and found that the bubble formation positions were nearly identical for a certain velocity. This made it possible to fit the data with a constant. The graph for airflow of 8.71 m/s is shown on the right in figure 2. The formation position remained nearly unchanged at $Z=0.052$ m. We then conducted experiments using the other four velocities. The five fitted position values for the five velocities were used to explore the relationship between breakup position and velocity ${v_0}$ - see figure 5. Using the measured values for ${v_N}$ and ${R_0}$, we can get ${v_0}$ using the flux conservation relation ${v_N}R_N^2 = {v_0}R_0^2$. In figure 5, we plotted $Z/R_0^{3/2}$ versus ${v_0}$ for the different experiments. The breakup position $Z$ reveals a nearly linear dependency on ${v_0}$, and the larger the velocity is, the further the breakup position moves away from the nozzle.

We then investigated the bubble formation process for the other three nozzles. The results were similar to that of the 0.75 cm nozzle. In figure 4, we use ${R_b}/{R_0}$ instead of ${R_b}$. The data for the four nozzles are distributed over a small range. ${R_b}/{R_0}$ is around two. We plotted $Z/R_0^{3/2}$ versus ${v_0}$ in figure 5. The reason why we define the quantity $Z/R_0^{3/2}$ will be given in the next section, and we find that all data sets lie on a single line. In the previous experiments, the Weber number $We = {\rho _b}v_N^2{D_N}/\sigma$, which corresponds to the upper airflow velocity ${v_{\max }}$, is about 289 $\sim$ 350 for the four nozzles. Here, ${D_N}$ is the diameter of the nozzle. The large Weber number means that the soap film could be destroyed, however, a detailed discussion of the Weber number is beyond the scope of the present work. For certain nozzle, the variation of the soap film column's diameters for different airflow are not obvious in experiment, so the influence of Bernoulli suction effect is negligible. The soap film generated by  plastic frame has wedge-like thickness variation in vertical direction, the film's thickness increases towards the bottom. The soap film is highly stretched when a soap film column is formed by the airflow, and the film's thickness is changed which causes variations of local surface tension on the soap film column based on the Gibbi-Marangoni effect.

\section{Theory}
In this section, we try to construct a simple physical model of our observations. We know that the pinch-off of thin water jets that emerges from taps is due to the Rayleigh-Plateau instability, which is driven by interfacial tension. The perturbation grows quickly and reduces the total surface area of the fluid thread by breaking it into tiny water droplets.

As found in the experiments (see figure 1), when a soap film is blown by nozzle, a film column is formed first. The jet can be regarded as circular, assuming that the initial perturbation generated at the film column is symmetric around the axle wire of the film column. This is well expressed using a cylindrical coordinate system. The perturbation of the film column radius, velocity, and the pressure on both sides of the film column can be expressed as:
\begin{eqnarray}
R = {R_0} + \varepsilon {e^{\omega t + ikz}}, \label{eq1}\\
\mathbf{v} = \mathbf{v_0} + \delta \mathbf{v},\label{eq2}\\
P = {P_0} + \delta P,\label{eq3}\\
P' = {P_0}^\prime  + \delta {P'_0}.\label{eq4}
\end{eqnarray}

Here, the perturbation amplitude $\varepsilon \ll {R_0}$, $\omega$ is the growth rate for the perturbation of the film column, and $k$ is the wave number of the perturbation in $z$-direction. $P$ is the pressure for the inner surface of the film column, $P'$ is the pressure on the outside of the film column. The perturbation velocity $\delta \mathbf{v}$ can be separated into a radial component $\delta {v_r} = V_r(r){e^{\omega t + ikz}}$ and an axial component $\delta {v_z} = V_z(r){e^{\omega t + ikz}}$, $V_r(r)$ and $V_z(r)$ are the amplitude of perturbation velocities of two components. We can also describe the perturbation on both sides of the film column using $\delta P = P(r){e^{\omega t + ikz}}$ and $\delta {P'_0} = P(r)'{e^{\omega t + ikz}}$. We substitute these perturbations into the Navier-Stokes equations and retaining terms to order $\varepsilon$. We then use the same procedure for the linearized continuity equation. The difference between the pressures at the surface of the film column is equal to the Laplace pressure, $\sigma \nabla  \cdot \mathbf{n} = P - P'$. For the inviscid limit, and the air density being equal outside and inside the column, the continuity of the displacements at the interface of film column results in the dispersion relation\cite{10}

\begin{equation}
\left[ {1 + \frac{{{K_0}(k{R_0}){I_1}(k{R_0})}}{{{K_1}(k{R_0}){I_0}(k{R_0})}}} \right]{\omega ^2} =  \frac{\sigma }{{\rho R_0^3}}[k{R_0} - {(k{R_0})^3}]\frac{{{I_1}(k{R_0})}}{{{I_0}(k{R_0})}}.
\end{equation}

Both ${I_i}$ and ${K_i}$ are modified Bessel functions of the first and second kind. A plot of the dispersion relation is shown in figure 3. $k{R_0} = 1$ is the cut-off wavenumber for $k{R_0} < 1$, whose wavelengths exceed the circumference of the film column, $2\pi {R_0} < \lambda$. This represents exponential growth of the perturbation and forces the film column into a bubble shape. The fastest growing mode occurs for $k{R_0} = 0.677$, i.e. $\omega _{\max }^2(\rho R_0^3/\sigma ) = 0.099$. We can estimate the characteristic breakup time $\tau  = 1/{\omega _{\max }}$ and get:

\begin{equation}
\tau  = 3.2\sqrt {\frac{{\rho R_0^3}}{\sigma }},
\end{equation}

The critical length range $L$ is the length range for which the film column begins to decompose into soap bubbles. It can be expressed as:

\begin{equation}
L = 3.2{v_0}\sqrt {\frac{{\rho R_0^3}}{\sigma }},
\end{equation}

We can approximately balance the capillary energy of a film column of length $L$ with that of the soap bubble, $(2\pi {R_0}L)\sigma  \simeq 4\pi R_b^2\sigma$, and $L$ is approximately equal to the wavelength, which is the fastest growing mode, $L =2\pi {R_0}/0.677$. This leads to an interesting conclusion:

\begin{equation}
\frac{{{R_b}}}{{{R_0}}} \simeq 2.15.
\end{equation}

The soap bubble radius is about 2.15 times greater than that of the film column radius. This ratio is independent of the surface tension, airflow, and soap film thickness.

\begin{figure}[h!]
\centering
\includegraphics[width=3.5in]{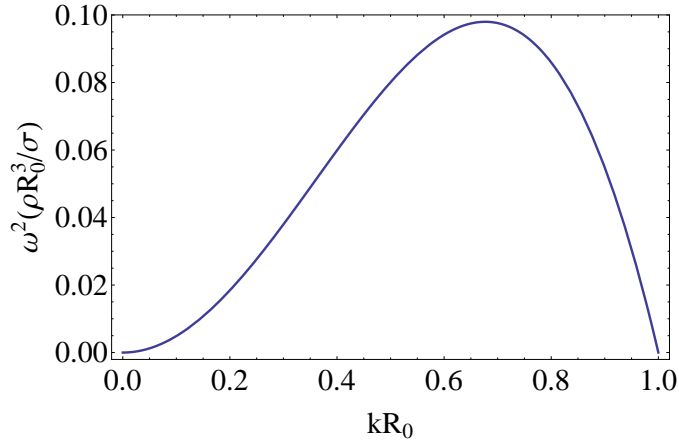}
\caption{Dimensionless growth rate ${\omega ^{\rm{2}}}\sqrt {\rho R_0^3/\sigma }$ as a function of dimensionless wave number $k{R_0}$ for the Rayleigh-Plateau instability.}
\label{fig4}
\end{figure}

We then study the formation position $Z$, which is the distance between the nozzle and the breakup point. We assume that the formation position $Z$ is equal  to the critical length range $L$, which is:

\begin{equation}
Z = L = 3.2{v_0}\sqrt {\frac{{\rho R_0^3}}{\sigma }}.
\end{equation}

For the Rayleigh-Plateau instability, an exponential growth of the initial perturbation occurs only for $k{R_0} < 1$. We use $\lambda  \sim v\sqrt {\frac{{\rho R_0^3}}{\sigma }}$, hence, the minimum velocity to appear instable is $v \sim \sqrt {\frac{\sigma }{{\rho {R_0}}}} $, which is consistent with the result given by Salkin et al.\cite{3}

\section{Results and discussion}
In figure 4, we show a comparison between the theoretical results using Eq.(8) for the bubble radius and the measured data. The prediction using a ratio of 2.15 is consistent with the experimental data. The bubble size is always 2.15 times that of the film column, and the ratio is independent of airflow, and surface tension coefficient. Figure 5 shows a comparison between the theoretical results using Eq.(9) for the bubble formation position and the measured data. Due to the different velocities there are different ${R_0}$ values, and we use $Z/R_0^{3/2}$ instead of $Z$.  Hence, Eq.(9) becomes $Z/R_0^{1.5} = (3.2\sqrt {\rho /\sigma }) {v_0}$. As the figure shows, the obtained data follow a single line. We plotted Eq.(9) in figure 5, which indicates that the theory matches the experimental data essentially. However, there are some deviations in figure 5. The evaporation of water in the soap film alters the surface tension coefficient, which could affect the result. Therefore, we choose the formation position $Z$ to be equal to the critical length range, in a first approximation.

\begin{figure}[h!]
\centering
\includegraphics[width=3.5in]{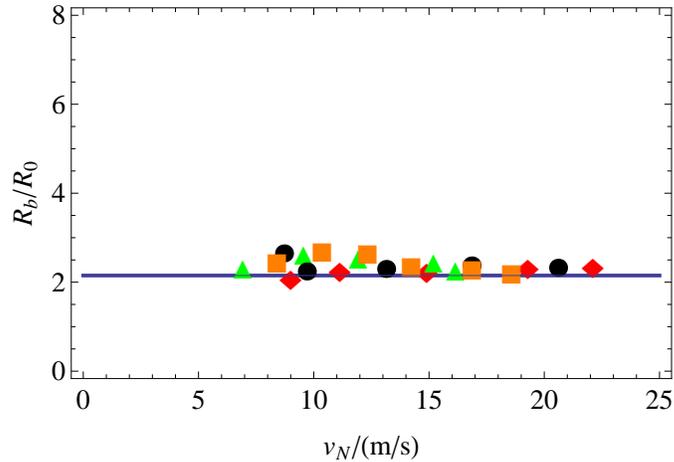}
\caption{Normalized plots ${R_b}/{R_0}$ versus air flow velocity ${v_N}$. Red rhombus is the data for nozzle 0.5 cm, black dots represent the data for nozzle 0.75 cm, orange triangle represent the data for nozzle 1.20,  and green square represent the data for nozzle 1.70 cm. The blue solid line indicates the ratio 2.15.}
\label{fig4}
\end{figure}

\begin{figure}[h!]
\centering
\includegraphics[width=3.5in]{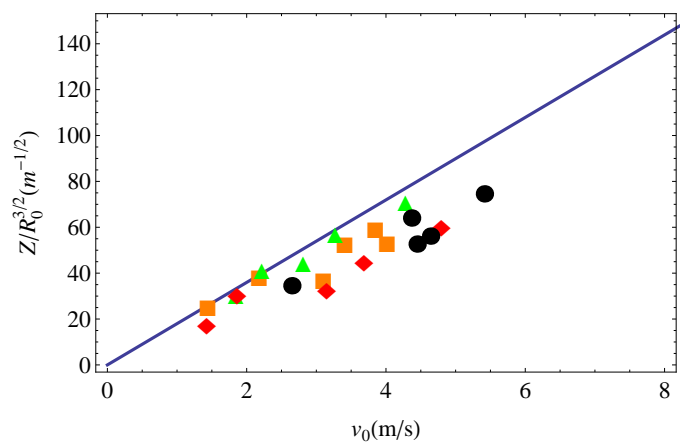}
\caption{Normalized plots $Z/R_0^{1.5}$ versus the velocity ${v_0}$. Red rhombus is the data for nozzle 0.5 cm, black dots represent the data for nozzle 0.75 cm, orange triangle represent the data for nozzle 1.20,  and green square represent the data for nozzle 1.70 cm. The blue solid line indicates the prediction of Eq.(9).}
\label{fig4}
\end{figure}

S.Leroux et.al \cite{11} studied liquid jets for different velocities, at low speed, the breakup length (the distance between the nozzle and the first disruption of the jet) increases linearly with velocity. Furthermore, the breakup length versus jet velocity should deviate from the linear relation for large velocities. This is consistent with our experiments at low velocity. However, in our experiment, the maximum airflow is about 25 m/s, and the larger airflow velocity causes to fracture the soap film immediately. Hence, the soap bubble cannot be blown with large velocities.

\section{Conclusions}
In this study, we investigated bubble size and formation position. We reach several interesting conclusions. We find that the ratio between soap bubble radius and soap film column remains essentially constant. Furthermore, the formation position increases linearly with the velocity. Rayleigh-Plateau instability theory can be used to explain our observations, and, based on some simple assumptions, we could derive equations for both bubble radius and its formation position. The theoretical results are consistent with the experimental data. However, the film thickness variations could affect the formation, we will investigate these effect in further work.

\ack
The authors gratefully acknowledge Dr.Yu Dai, Pro.Jun Zou for their help.

\end{document}